# STANDARDIZED MEDICAL IMAGE CLASSIFICATION ACROSS MEDICAL DISCIPLINES

*Simone Mayer[1], Dominik Müller[*,1,2] and Frank Kramer[1]*

[1]IT-Infrastructure for Translational Medical Research, University of Augsburg, Germany
[2]Medical Data Integration Center, Institute for Digital Medicine, University Hospital Augsburg, Germany

## ABSTRACT

**AUCMEDI is a Python-based framework for medical image classification. In this paper, we evaluate the capabilities of AUCMEDI, by applying it to multiple datasets. Datasets were specifically chosen to cover a variety of medical disciplines and imaging modalities. We designed a simple pipeline using Jupyter notebooks and applied it to all datasets. Results show that AUCMEDI was able to train a model with accurate classification capabilities for each dataset: Averaged AUC per dataset range between 0.82 and 1.0, averaged F1 scores range between 0.61 and 1.0. With its high adaptability and strong performance, AUCMEDI proves to be a powerful instrument to build widely applicable neural networks. The notebooks serve as application examples for AUCMEDI.**

## INTRODUCTION

Deep neural networks are a promising tool for medical image classification. Especially convolutional neural networks (CNN) can excel in classification tasks involving images. By now, their performances are able to rival human experts [1,2].

Including deep neural networks in clinical practice remains challenging for several reasons: Collecting and preparing data for a neural network requires both medical as well as technical expertise. Improper preprocessing may cause a bias in the dataset which could ultimately impact the generalization performance [3]. Furthermore, physicians, which are responsible for patient care, are often reluctant to accept a machine-made decision, if the decision is not transparent or comprehensible [4]. Even reliable results generated by a well-performing "black-box" can lead to mistrust in continuous reliability [3]. Finally, solving a classification problem with a neural network model often results in complex, isolated solutions. Due to their lack of modularity, such implementations are limited in reusability or even reproducibility.

The proposed framework AUCMEDI aims to simplify such implementations. It allows simple usability resulting in requiring only little domain knowledge and merely basic programming skills. Furthermore, the framework features modularity with regard to network architectures, datasets, and even medical disciplines. The purpose of this study is to demonstrate AUCMEDI's capabilities in terms of standardized application and adaptability. To evaluate the framework, a single pipeline using AUCMEDI was designed and applied to a wide range of image classification tasks from diverse medical disciplines.

## METHODS

### Datasets

In order to demonstrate wide applicability, the AUCMEDI framework was applied to datasets from the following medical disciplines and procedures: Dermatology via dermatoscopy, gastroenterology via endoscopy, histopathology via microscopy, neurology via magnetic resonance imaging, radiology via X-ray as well as computed tomography, gynecology via ultrasound, and ophthalmology via retinal imaging. For this experiment, only 2D imaging was considered for classification. A brief overview of the analyzed datasets can be found in Table 1.

### Dataset Descriptions

The dermatology dataset providing dermatoscopy images was published by the International Skin Imaging Collaboration (ISIC) [5–7] and consists of the following classes: Melanoma, melanocytic nevus, basal cell carcinoma, actinic keratosis, benign keratosis, dermatofibroma, vascular lesion, squamous cell carcinoma, and unknown. Eight of these are present in the training data used for this evaluation. The unknown class does not occur and was explicitly excluded during preparations. The gastroenterology dataset providing endoscopy images was published by Borgli et al. [8]. Two subsets from the dataset were used for separate anatomical landmark classification tasks, one for the upper gastrointestinal (GI) tract and one for the lower

**Table 1:** Overview of the datasets utilized in the broad application experiments.

| Medical Discipline | Dermatology | Gastroenterology | Histopathology | Neurology | Radiology | Radiology | Gynecology | Ophthalmology |
|---|---|---|---|---|---|---|---|---|
| Medical Procedure | Dermatoscopy | Endoscopy | Microscopy | MRI | X-ray | CT | Ultrasound | Retinal Imaging |
| Task | Skin Lesion Classification (Melanoma) | Landmark Classification (upper & lower) | Invasive Ductal Carcinoma Detection | Brain Tumor Classification | Pneumonia Detection | COVID-19 Classification | Breast Cancer Screening | Retinal Multi-Disease Detection |
| Reference | [5–7] | [8] | [9,10] | [11] | [12] | [13] | [14] | [15] |
| Samples | 25,331 | 2,695 (upper) 1,409 (lower) | 277,524 | 7,022 | 5,856 | 19,685 | 1,578 | 3,588 |
| Classes | 8 | 3 (upper) 3 (lower) | 2 | 4 | 2 | 3 | 3 | 28 |
| Multi-label | No | No | No | No | No | No | No | Yes |

GI tract. The upper GI tract dataset consists of the following classes: Pylorus, retroflex-stomach, and z-line. Accordingly, the lower GI tract dataset consists of the following classes: Cecum, ileum, and retroflex-rectum. The histopathology dataset providing microscopy images was published by Madabhushi et al. [9,10] and consists of the following classes based on invasive ductal carcinoma presence: Negative and positive. Images were originally mapped to certain patients. This information was discarded and not respected for splitting the dataset. The neurology dataset providing 2D magnetic resonance imaging (MRI) slides was published by Msoud Nickparvar [11] and consists of the following classes: Glioma, meningioma, pituitary, and no tumor presence. The radiology dataset providing 2D computed tomography (CT) slides was published by Ning et al. [12] and consists of the following classes: Non-informative finding in which lung parenchyma was not captured for any judgment, positive finding in which features associated with COVID-19 pneumonia could be unambiguously discerned, and negative finding in which features in both lungs were irrelevant to COVID-19 pneumonia. The dataset consists of original and preprocessed CT scans. Only original images were considered. The radiology dataset providing X-ray images was published by Kermany et al. [13] and consists of the following classes depending on pneumonia presence: Negative and Positive. The gynecology dataset providing ultrasound images was published by Al-Dhabyani et al. [14] and consists of the following classes: Normal, benign, and malignant. The ophthalmology dataset providing retinal images was published by Pachade et al. [15] and consists of 28 multi-label classes for various medical conditions.

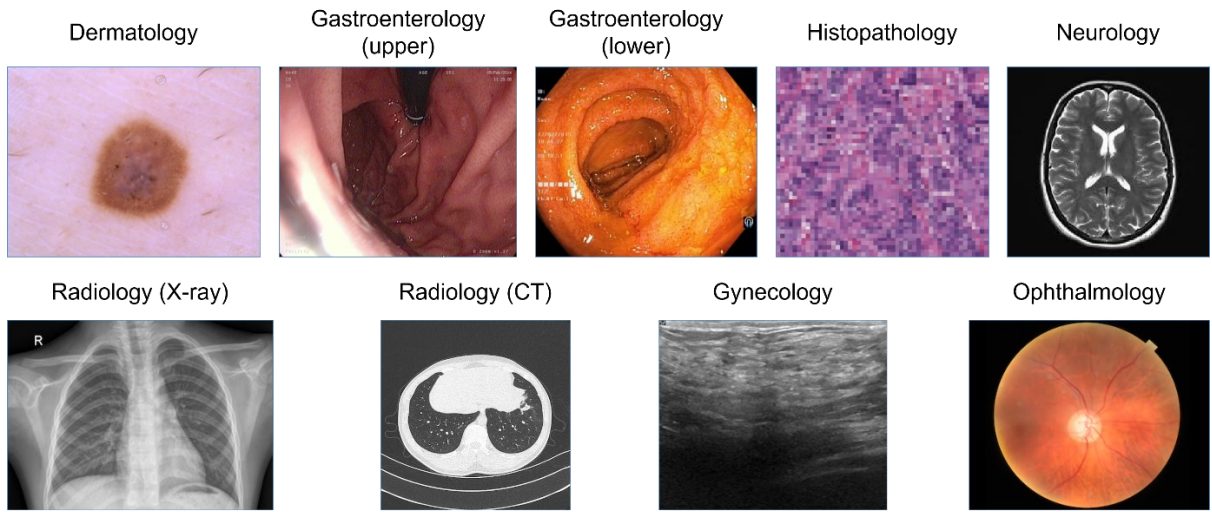

**Figure 1:** Exemplary samples of the analyzed datasets.

## Pipeline Design

AUCMEDI applies several preprocessing techniques by default: Images are augmented by flipping and rotating, as well as applying changes to brightness, contrast, saturation, hue, and scale. Afterward, images are resized to 224x224 pixels and standardized according to Z-score.

Datasets are split into three sets: 70% of the data are used for training, 10% are used as validation during training, and 20% are used as hold-out set for testing. The split is stratified to keep the proportional class distribution similar in all sets.

The neural network model is the core of the AUCMEDI pipeline. It consists of one model based on the DenseNet121 architecture [16]. To keep the required training time at a minimum as well as improve overall generalizability, a transfer learning approach was applied: AUCMEDI internally uses the defined architecture's pretrained weights and fine-tunes these weights using the provided training data. The pretrained weights were based on ImageNet [17]. Training is done for a maximum of 500 epochs. To further reduce the required training time, an early stopping strategy was employed. Additionally, the learning rate was dynamically adjusted to improve the learning progress. Based on the hold-out set for which predictions were generated with the fitted model, the classification performance was evaluated. For performance assessment, the following metrics were computed: Accuracy, F1-score, the ROC curve, and area under the ROC curve (AUC).

The ability to understand a classification decision by a model, commonly known as explainability, is fundamental in the medical context. One possibility to explain a machine-made decision is by highlighting relevant regions within the classified image which can be achieved by a heatmap visualization. For multi-class datasets, one sample per class was evaluated utilizing the Grad-CAM algorithm [18], whereas, in the case of a multi-label task, Grad-CAM images for only one sample from the hold-out set for all possible classes were generated.

## RESULTS AND DISCUSSION

Training for each of the 9 experiments was performed on a single NVIDIA TITAN RTX GPU. Due to the early stopping strategy and deployed simplistic architecture in terms of the number of parameters, training was finished after on average 51.6 epochs.

The classification performance for all experiments revealed strong predictive capabilities. Achieved AUC, Accuracy, and F1-scores are summarized in Figure 1. In the following sections, noteworthy observations for each dataset are elaborated and resulting insights are discussed.

## Individual Dataset Results

The dermatology dataset containing images of skin lesions for melanoma classification revealed a strong class imbalance. Nevertheless, the applied model was able to achieve accurate classification performance with an F1-score of approximately 0.6 and an AUC of more than 0.9. The two subsets from the gastroenterology dataset were evaluated as isolated landmark classification tasks. The computed results showed a discrepancy in their performances. By comparing the class distributions between the two datasets, the massive class imbalance for the lower GI tract can be noticed. The ileum class consists of nine images, as opposed to the cecum class with more than 1,000 samples. Further detailed analysis of the classification results confirmed that inferior predictions are associated with the underrepresented ileum class. The histopathology dataset is a binary classification task to identify invasive ductal carcinoma, the most common subtype of breast cancer [9]. Despite the provided small image sizes of only 50x50 pixels, the model was able to reliably distinguish healthy from pathological slides. Since the patient information was omitted during preprocessing, it would be promising to include the available metadata through the metadata interface of AUCMEDI in this only image-based approach. The neurology dataset represents a large as well as class-balanced set of images. This is why the model was able to achieve an exceptional classification performance of an F1-score of 0.9 and an AUC of 1.0. In the CT-based radiology dataset, the model achieves remarkable classification performance on the task to detect COVID-19 infected regions. The task of the X-ray-based radiology dataset was to distinguish if a patient has pneumonia or not. The calculated F1-score of 0.95 and AUC of 1.0 revealed that the fitted model was able to achieve powerful classification performance, as well. However, a slight class imbalance was also observed in the gynecology dataset, where breast cancer was to be identified using ultrasound images. The classification performance on this dataset was adequate but favored classifying samples as normal, a class which contains merely 266 images, and was cautious in assigning to samples the class benign, which was with 891 samples the most common class in the gynecology dataset. A more balanced dataset or a more complex neural network architecture might improve predictive capabilities. The ophthalmology dataset is the only multi-label task that was evaluated in this study. Due to the complexity of this task, the averaged AUC was approximately 0.8 and notably lower than other datasets, but still strong considering the challenging character of this dataset [15]. Due to the multi-label context, the model did not take one most likely label to assign (softmax activation) but weights each label with a probability for the given sample (sigmoid activation). An experimental analysis of non-optimal predictions also revealed that the fitted model was often capable of correctly identifying other present diseases in the sample demonstrating high sensitivity.

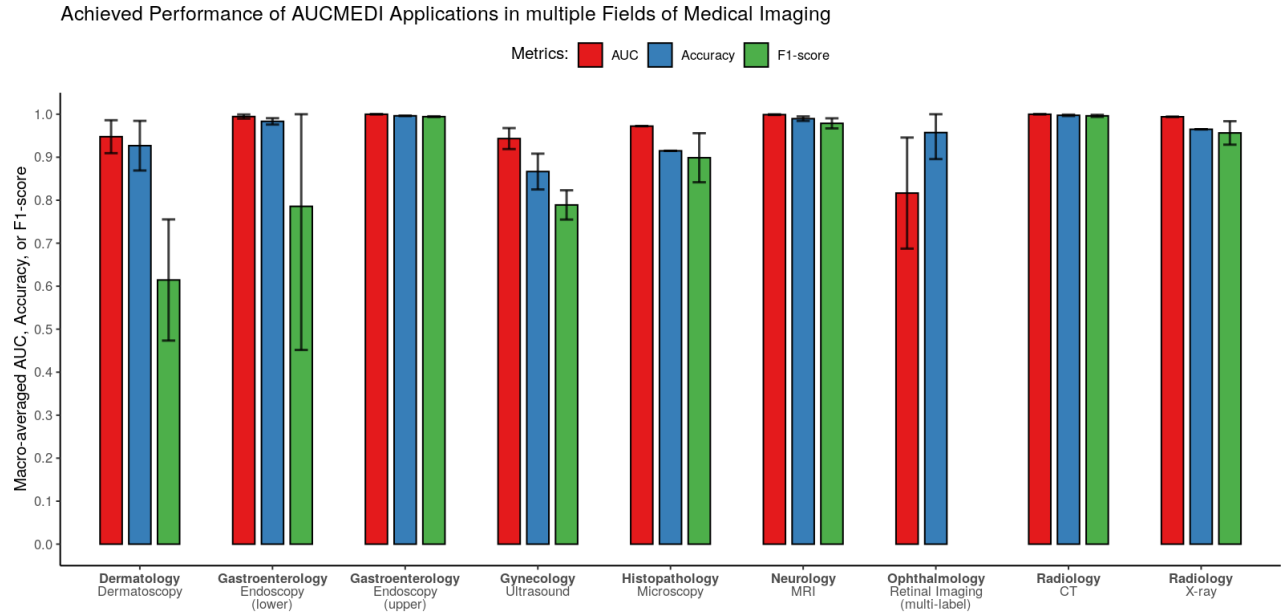

**Figure 1:** Summary of the achieved performance in the conducted experiments. The bar plots represent the average performance by mean and the error bars represent the standard deviation. Since the ophthalmology dataset is a multi-label task, F1-scores hold limited informative value and were omitted.

### Performance and Adaptability

We concluded that with an averaged AUC between 0.8 and 1.0, all fitted models showed a strong performance. That underlines the potential of AUCMEDI: The standardized pipeline was successfully applied to a diverse range of modalities which is highly promising for clinical application. This demonstrates that there is no need for dataset-specific implementations or distinctions between binary, multi-class, and multi-label problems. Instead, clinical researchers are able to design a state-of-the-art pipeline without extensive domain knowledge, apply it to the classification task, and achieve accurate predictions. AUCMEDI's high-level programming interface is simple to use even with limited deep learning knowledge or programming experience.

### Class Imbalance Influence on Performance

Detailed performance metrics for each dataset revealed, that a well-balanced, large dataset is more likely to generate accurate classifications. Evaluating the two gastroenterology datasets, both subsets had the task to assign anatomical landmarks to images and both contained three distinct classes that were to be identified by the machine. However, the subset relating to the lower GI tract was massively unbalanced. The ileum class contains less than 10 samples, whereas the cecum class contains more than 1,000 samples. In contrast, the upper GI tract subset was more balanced. The stratified sampling split used for the pipeline reflected these proportions. The computed AUC, Accuracy, and F1-scores for both subsets indicated that the landmark classification could work reliably but requires a balanced dataset. A similar phenomenon was observed for the ophthalmology dataset. Its class distribution showed several classes with less than 10 occurrences which had also the lowest performance scores.

## Importance, Potential, and Challenges of XAI

Due to the sparsity and diversity of medical data, the capability to understand a machine-made decision is crucial in the medical domain. Particularly promising are insights into images from a modality that displays medical instrumentation, for instance X-rays. The processed X-ray dataset in this study, for which a sample is visualized in Figure 3 (parts A and B), clearly shows electrodes and what appears to be a drainage. The model identified this sample as having pneumonia based on these instruments and not based on lung tissue. This example illustrates the importance of explainability in the medical context. Furthermore, it is important to note that XAI is not limited to explaining a decision of a model, but can hint humans to image sections, they might not even consider to be relevant for the condition. The dermatoscopic images of suspicious-looking skin patches is an adequate example which is illustrated in Figure 3 (parts C and D). The figure shows a squamous cell carcinoma and its respective Grad-CAM visualization. The XAI image marks a small section of skin just below the abnormal tissue region which apparently revealed information about the condition. However, the gold standard to diagnose this type of cancer is an invasive biopsy and the subsequent histopathological examination [19]. With the support of XAI technologies, it may be possible for medical experts to develop more non-invasive diagnostic measures. Nevertheless, XAI still faces various challenges: The smallest cluster of pixels can be enough for a neural network model to make a decision with high confidence. But if the image has a low resolution and details are humanly not possible to identify, XAI would hardly yield any benefits.

## CONCLUSIONS

In this study, the capabilities of the proposed framework AUCMEDI were evaluated. The pipeline for medical image classification was developed and applied to 8 datasets from a wide range of medical imaging modalities. Each of the trained models achieved an exceptional classification performance. This evaluation illustrates that AUCMEDI is a powerful tool to build modular and reusable classification pipelines without restrictions to specific datasets or medical disciplines. However, the application of a single pipeline to several datasets also emphasizes the relevance of data. Detailed analysis revealed that the dips in classification performance throughout the experiments were caused by a significant class imbalance within the respective dataset.

That a standardized pipeline applied to diverse classification tasks was able to generate independently well-performing models, is promising. As future work, we plan to validate AUCMEDI in clinical studies to evaluate the framework's impact and usability in clinical environments.

## DECLARATIONS

### Availability of data and materials

To ensure full reproducibility, the complete code of this study is available in the AUCMEDI Git repository and accessible in the example registry in the AUCMEDI documentation: https://frankkramer-lab.github.io/aucmedi/examples/framework/.

### Competing interests

Simone Mayer and Dominik Müller are contributors to AUCMEDI.

### Funding

This work is a part of the DIFUTURE project funded by the German Ministry of Education and Research (Bundesministerium für Bildung und Forschung, BMBF) grant FKZ01ZZ1804E.

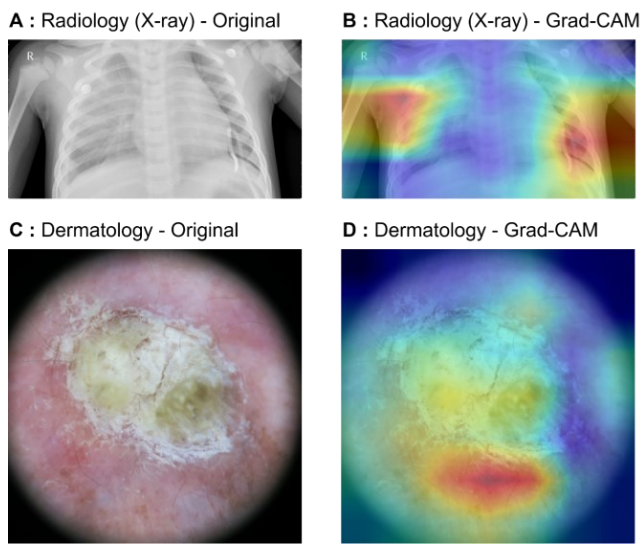

**Figure 3:** XAI visualization of samples from the radiology (X-ray) and dermatology dataset.